# Solar Sails, Optical Tweezers, and Other Light-Driven Machines

Masud Mansuripur

College of Optical Sciences, The University of Arizona, Tucson, Arizona 85721

**Abstract**. Electromagnetic waves carry energy, linear momentum, and angular momentum. When light (or other electromagnetic radiation) interacts with material media, both energy and momentum are usually exchanged. The force and torque experienced by material bodies in their interactions with the electromagnetic field are such that the energy as well as the linear and angular momenta of the overall system (i.e., the system of field plus matter) are conserved. Radiation forces are now used routinely to trap and manipulate small objects such as glass or plastic micro-beads and biological cells, to drive micro- and nano-machines, and to contemplate interstellar travel with the aid of solar sails. We discuss the properties of the electromagnetic field that enable such wide-ranging applications.

**1. Introduction**. Radiation pressure is a direct manifestation of the momentum carried by electromagnetic waves. First suggested by Johannes Kepler in his treatise *De Cometis*, radiation pressure is (partially) responsible for the tails of the comets pointing away from the Sun. According to this hypothesis, the solar ray pressure is responsible for the deflection of the comet tails; see Fig. 1. Although the observed deflections could not be explained *solely* on the basis of light pressure, Kepler's hypothesis played a significant role in understanding the effects of light pressure in the universe. The force and torque exerted by laser beams are now used routinely in laboratories to trap and manipulate tiny objects, such as biological cells, and to drive/control small functional devices such as micro- and nano-motors.

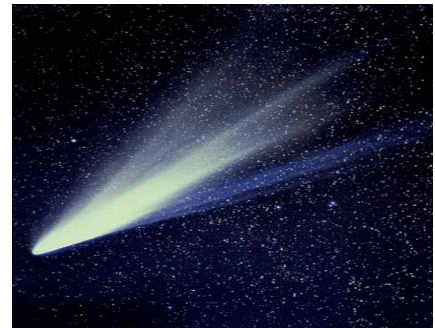

**Fig. 1**. Comet West photographed in 1976. The white tail is produced by evaporated dust and debris that are pushed away from the Sun by the solar radiation pressure.

Perhaps the simplest way to deduce from first principles that electromagnetic (EM) fields carry momentum is by means of the Einstein box thought experiment [1]. Shown in Fig. 2 is an empty box of length $L$ and mass $M$, placed on a frictionless rail, and free to move forward or backward. At some point in time, a blob of material attached to the left wall emits an EM pulse of energy $\mathcal{E}$ and momentum $\boldsymbol{p}$, which travels the length of the box and gets absorbed by another blob attached to the wall on the right-hand side. The recoil velocity of the box is thus $-\boldsymbol{p}/M$, the time of flight is $L/c$, and the box displacement along the rail is $-(\boldsymbol{p}/M)(L/c)$. Associating a mass $m = \mathcal{E}/c^2$ with the EM pulse and assuming that $M \gg m$, it is easy to see that the displacement of the center-of-mass of the system is proportional to $(\mathcal{E}/c^2)L - M(\boldsymbol{p}/M)(L/c)$. In the absence of external forces acting on the box, however, its center-of-mass is not allowed to move. Setting the net displacement equal to zero, we find $p = \mathcal{E}/c$. Thus, in free space, a light pulse of energy $\mathcal{E}$ carries a momentum $p = \mathcal{E}/c$ along its direction of propagation.

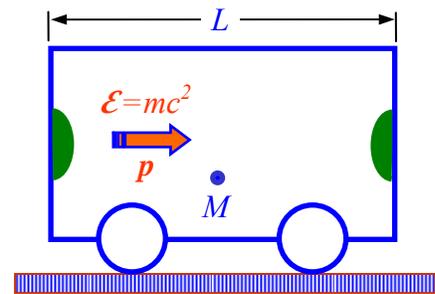

**Fig. 2**. Einstein box gedanken experiment.

A straightforward application of radiation pressure is found in the concept of solar sails; see Fig. 3. At 1.0 astronomical unit (i.e., the Sun-Earth distance), sunlight provides around 1.4 kw/m² of

EM power density. Dividing this by the speed of light $c$ and multiplying by 2 (to account for momentum reversal upon reflection from the sail) yields a pressure of $\sim 9.4\,\mu\text{N/m}^2$. Over a sufficiently long period of time, the continuous force of sunlight exerted on a large-area solar sail can propel a small spacecraft to speeds comparable to or greater than those achievable by conventional rockets.

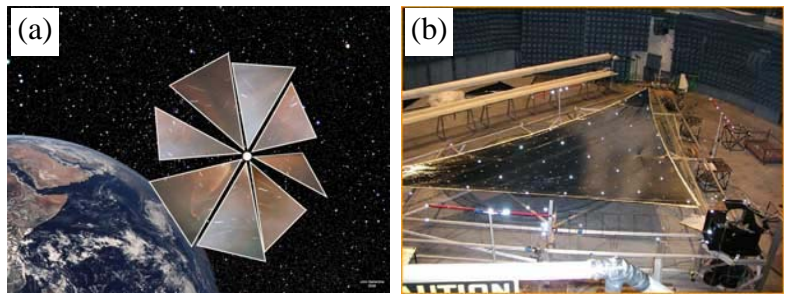

**Fig. 3**. (a) Artist's conception of a small solar-sail-driven spacecraft traveling away from the sun outside the Earth's atmosphere. (b) A 10-meter solar sail sits fully deployed in a vacuum chamber at NASA's Langley Research Center.

**2. Reflection of light from a perfect mirror**. One way to analyze the effect of reflection from a mirror is by considering a pulse of light containing $N$ photons of frequency $\omega$ arriving at the surface of a perfect reflector, as shown in Fig. 4. Before reflection, the light pulse has energy $\mathcal{E} = N\hbar\omega$ and momentum $p = N\hbar\omega/c$, while the mirror's kinetic energy and momentum are both equal to zero. After reflection, the pulse, now Doppler-shifted to a lower frequency $\omega'$, will have energy $\mathcal{E} = N\hbar\omega'$ and momentum $p = -N\hbar\omega'/c$, while the mirror's kinetic energy and momentum will be $\tfrac{1}{2}M_o V^2$ and $M_o V$, respectively. Solving the energy and momentum conservation equations yields the unknown parameters $V$ and $\omega'$, as follows:

$$V/c = \sqrt{1 + \frac{4N\hbar\omega}{M_o c^2}} - 1 \approx \frac{2N\hbar\omega}{M_o c^2}\left(1 - \frac{N\hbar\omega}{M_o c^2}\right), \qquad (1)$$

$$\omega' = \frac{M_o c^2}{N\hbar}\left(\sqrt{1 + \frac{4N\hbar\omega}{M_o c^2}} - 1\right) - \omega \approx \omega\left(1 - \frac{2N\hbar\omega}{M_o c^2}\right). \qquad (2)$$

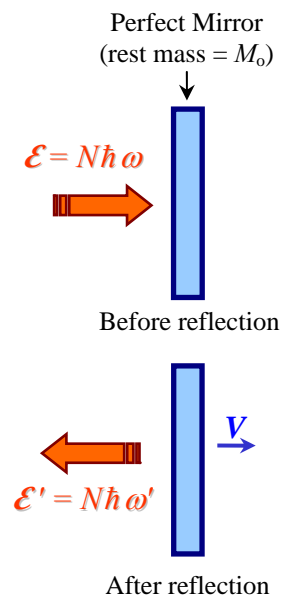

**Fig. 4**. Reflection of a light pulse from a perfect mirror.

The approximate expressions on the right-hand sides of the above equations are obtained by assuming that the pulse energy $N\hbar\omega$ is substantially less than the mirror's rest energy $M_o c^2$. In this regime, $M_o V \approx 2N\hbar\omega/c$ and $\omega' \approx \omega(1 - V/c)$. We mention in passing that using the relativistic expressions $M_o c^2/\sqrt{1 - V^2/c^2}$ and $M_o V/\sqrt{1 - V^2/c^2}$ for the mirror's energy and momentum in the balance equations would yield a fully relativistic formula for the Doppler shift $\Delta\omega = \omega' - \omega$.

**3. Radiation pressure and the Lorentz force**. The Lorentz force $f$ on a point-charge $q$, moving with velocity $V$ in a region of space containing the EM fields $(E, B)$ is given by $f = q(E + V \times B)$; see Fig. 5(a). This leads to the force density equation $F(r,t) = \rho(r,t)E(r,t) + J(r,t) \times B(r,t)$, where $\rho$ and $J$ are the electric charge and current densities, respectively [2]. In vacuum, and also in regions of space devoid of polarization $P$ and magnetization $M$, we have $B = \mu_o H$ and $D = \varepsilon_o E$, where $\mu_o$ and $\varepsilon_o$ are the permeability and permittivity of free space. Figure 5(b) shows a perfectly conducting mirror illuminated by a plane, monochromatic wave at normal incidence. The ratio of the electric field $E$ to the magnetic field $H$ of the plane-wave in vacuum is $E/H = Z_o = \sqrt{\mu_o/\varepsilon_o}$, where $Z_o$ is the impedance of free space. At the mirror surface the incident and reflected $E$-fields cancel out, but the $H$-fields add up to a strength of $2H$. Maxwell's boundary conditions then dictate that the surface current $J_s$ must be equal to $2H$ in the direction perpendicular to the $H$-field immediately above the surface. The time-averaged Lorentz force per unit surface area is thus given by $½J_s \times B = \mu_o H^2 = \varepsilon_o E^2$. This is consistent with the incident and reflected beams each having a momentum density $p = S/c^2$, where $S = ½E \times H$ is the time-averaged Poynting vector and $c = \sqrt{\mu_o \varepsilon_o}$ is the speed of light in vacuum [3]. For an incident optical power of 1.0 W/mm², therefore, the radiation pressure on the mirror surface will be 6.67 nN/mm².

When a beam of light is reflected from a solid mirror, the mirror does *not* move forward as a solid block; this would violate special relativity as the "news" of action at the front facet will have instantaneously reached the rear facet. Rather, an elastic wave is launched at the front facet, which proceeds to propagate within the body of the mirror, pushing the molecules forward in an orderly fashion [4]. The schematic diagram in Fig. 6 shows how molecular layers push and pull each other in an effort to distribute the momentum received from the impact of the light pulse. These vibrations eventually settle down and the mirror moves forward at a uniform velocity.

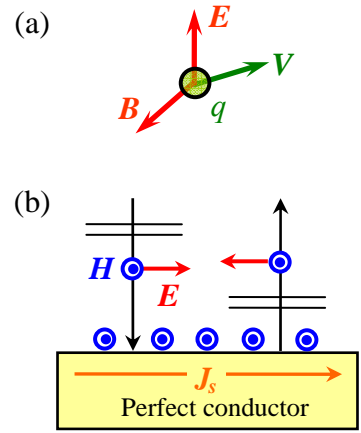

**Fig. 5**. (a) The Lorentz law expresses the force of the $E$ and $B$ fields on the point-charge $q$ moving with velocity $V$. (b) A linearly-polarized plane wave is normally incident on a perfect conductor. While the $E$-field at the conductor surface vanishes, the $H$-field, being the sum of the incident and reflected waves, exerts a Lorentz force on the induced surface current $J_s$.

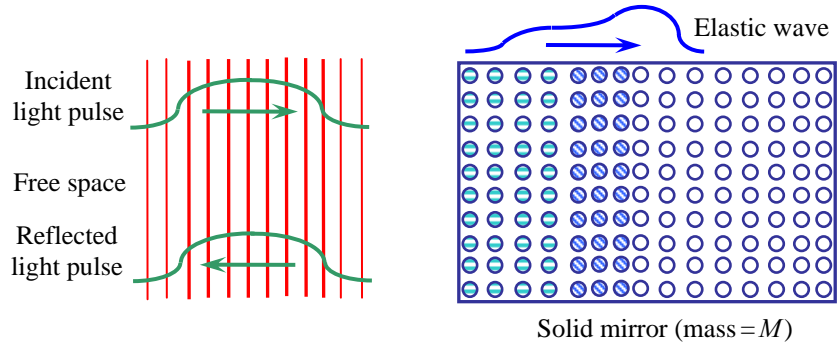

**Fig. 6**. Reflection of a light pulse from the front facet sets up an elastic wave within the body of the mirror.

**4. Radiation pressure on a dielectric wedge**. Figure 7 shows a dielectric wedge of refractive index $n$, illuminated by a collimated beam of $p$-polarized light at Brewster's angle of incidence ($\tan\theta_B = n$). At this incidence angle, the reflectance of the surface for $p$-light is exactly zero. The apex angle $\phi$ of the wedge is chosen to allow the beam to exit the prism also at Brewster's angle. Thus, with no reflections at entrance and exit facets, the entire incident beam turns around and leaves the wedge, as shown. The action of the prism on the beam has thus changed the direction of its momentum. The net change in the optical momentum upon transmission must therefore be transferred to the dielectric medium as a force ***F*** that would propel the wedge forward. An alternative determination of the force ***F*** involves calculating bound-charge and bound-current distributions inside the prism and at its surfaces, followed by a calculation of the Lorentz force of the *E*- and *H*-fields on these charges and currents. The two methods of analysis outlined above yield identical results, thus affirming the consistency of the force law with momentum conservation [5].

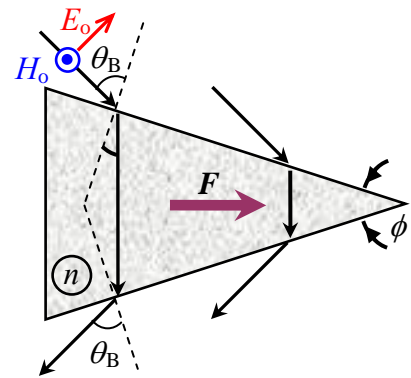

**Fig. 7**. Deflection of a beam by a glass prism.

**5. Optical tweezers**. The first optical traps were built by Arthur Ashkin at AT&T Bell laboratories in the 1970s. "Levitation traps" used the upward-pointing radiation pressure to balance the downward pull of gravity, whereas "two-beam traps" relied on counter-propagating beams to trap particles [6,7]. Then, in 1986, Ashkin and colleagues realized that the gradient force alone would be sufficient to trap small particles. They used a single tightly focused laser beam to trap a transparent particle in three dimensions [8,9]. The principle of single-beam trapping is shown in Fig. 8. A small spherical dielectric bead of refractive index $n_{bead}$ is immersed in some liquid of refractive index $n_{liquid}$. A laser beam is focused from above into the glass bead, with the focal point placed slightly above the center of the sphere. (Only two of the incident rays are shown, but the remaining rays behave essentially in the same way.) The bending of the rays by the glass bead causes them to exit with a smaller deviation from the optical axis. The projection of the exiting rays' momenta on the optical axis is thus greater than that of the incident rays. Stated differently, optical momentum along the *z*-axis increases upon transmission through the bead. In the process, this change of optical momentum is transferred as a lift force to the glass bead, helping to support it against the pull of gravity. Additionally, it is not difficult to show that, if the bead is laterally displaced from equilibrium, the resulting gradient force will return it to its original position; in other words, the equilibrium is a stable one.

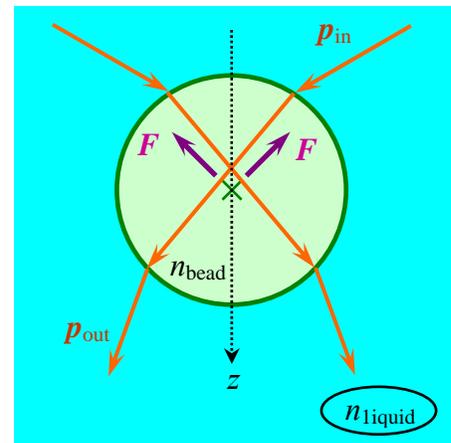

**Fig. 8**. Single-beam optical trap.

**6. Electromagnetic spin and orbital angular momenta**. As mentioned earlier, the linear momentum density (i.e., momentum per unit volume) of an electromagnetic field is given by $\boldsymbol{p}(\boldsymbol{r},t) = \boldsymbol{S}(\boldsymbol{r},t)/c^2$, where ***S*** is the Poynting vector and *c* is the speed of light in vacuum. The angular momentum density with respect to the origin of coordinates is thus given by $\boldsymbol{\mathcal{J}}(\boldsymbol{r},t) = \boldsymbol{r} \times \boldsymbol{S}(\boldsymbol{r},t)/c^2$. A pulse of light having a finite duration and occupying a finite volume in space will carry, at any given time, a certain amount of linear and angular momenta, which amounts can be determined by integrating the corresponding momentum densities over the region of space occupied by the pulse at any given instant of time. In the absence of interactions with material media (i.e., when the pulse resides in free space), one can show, using Maxwell's equations, that the total linear momentum and also the total angular momentum of a given pulse remain constant in time, i.e., the linear and angular momenta of the pulse are conserved [10]. If the light pulse enters a region of space where material media reside, it will exert forces and torques on various parts of these media in accordance with a generalized form of the Lorentz force law [11,12]. Such exchanges between fields and material media cause the EM momenta (linear as well as angular) to vary in time. These variations, however, are always accompanied by corresponding variations in the linear and angular momenta of the material media (mechanical momenta), in such a way as to conserve the total momentum of the system of fields plus media, be it linear or angular, at all times [13-19].

The angular momentum of a light pulse in free space could arise as a trivial consequence of its center-of-mass trajectory (a straight-line along the linear momentum of the pulse) not going through the chosen reference point. Selecting a reference point on the center-of-mass trajectory then eliminates this trivial (extrinsic) contribution. The remaining contributions to angular momentum can be divided into two categories: spin and orbital angular momenta [20-25]. Roughly speaking, spin has to do with the degree of circular polarization of the light pulse, whereas orbital angular momentum arises from spatial non-uniformities of amplitude and phase that render the beam asymmetric. Vorticity, which is associated with a continuous increase or decrease of phase around closed loops in the beam's cross-sectional plane, is a particularly interesting (and useful) source of orbital angular momentum [26].

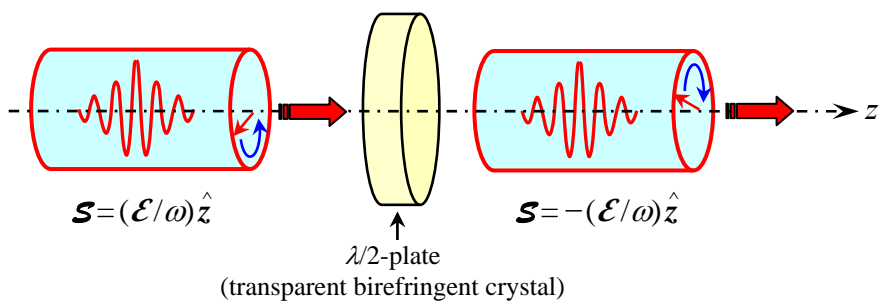

**Fig. 9**. A circularly-polarized light pulse of energy $\mathcal{E}$ and frequency $\omega$ carries a spin angular momentum $\boldsymbol{S} = \pm(\mathcal{E}/\omega)\hat{z}$. Upon transmission through a half-wave plate, the change in the optical angular momentum ($2\boldsymbol{S}$) is transferred to the wave-plate, thereby setting the plate spinning around the $z$-axis.

A circularly-polarized light pulse of energy $\mathcal{E}$ propagating along the $z$-axis carries a spin angular momentum $\boldsymbol{S} = \pm(\mathcal{E}/\omega)\hat{z}$ [27]. The $\pm$ signs indicate the dependence of the direction of $\boldsymbol{S}$ on the handedness of circular polarization (i.e., right or left). Such a light pulse, upon passing through a half-wave plate, will have its sense of polarization and, consequently, its direction of spin angular momentum (SAM) reversed. Conservation of angular momentum then requires the transfer of $2\boldsymbol{S}$ units of angular momentum to the half-wave plate, as shown in Fig. 9. The passage of the light pulse thus sets the wave-plate spinning around the $z$-axis, a phenomenon that has been exploited in optically driven micro-machines [28, 29].

When a collimated beam of light passes through a transparent spiral ramp, as depicted in Fig. 10, the emergent beam acquires optical vorticity, which carries orbital angular momentum (OAM). Once again, conservation of angular momentum requires the transfer of an equal but opposite angular momentum to the spiral ramp. Both SAM and OAM may be used to drive micro-machines. While the SAM associated with a single photon can only have a magnitude of $\pm\hbar$ (reduced Planck's constant), the magnitude of OAM could be any integer multiple of $\hbar$.

**7. The Abraham-Minkowski Controversy**. Whereas the linear momentum associated with a photon of frequency $\omega$ in vacuum has the well-known value of $\hbar\omega/c$, its strength inside a dielectric medium has been debated for over a century [30-32]. According to H. Minkowski's theory, the photon momentum inside a dielectric of refractive index $n$ is $n\hbar\omega/c$, while M. Abraham's analysis yields the vastly different value of $\hbar\omega/nc$. To resolve the controversy, we believe it is best to examine the foundations of the classical Maxwell-Lorentz theory of electrodynamics.

In the classical theory of electrodynamics, Maxwell's macroscopic equations describe the relation between the fields ($\boldsymbol{E}, \boldsymbol{D}, \boldsymbol{H}, \boldsymbol{B}$) and their sources (electric charge- and current-densities, $\rho_{\text{free}}$ and $\boldsymbol{J}_{\text{free}}$, as well as polarization $\boldsymbol{P}$ and magnetization $\boldsymbol{M}$). The relationship between EM energy and the fields is not directly and unambiguously discernible from Maxwell's equations, but if one accepts the hypothesis that the Poynting vector $S(r,t) = E(r,t) \times H(r,t)$ represents the rate of flow of EM energy per unit area per unit time, then all questions about EM energy can be answered within the context of Maxwell's macroscopic equations. A similar ambiguity exists with regard to the momentum density of EM fields: while Abraham arrives at the expression $p_A = E \times H/c^2$ for the EM momentum density, the corresponding expression based on Minkowski's approach is $p_M = D \times B$. Of course, in free space, where $D = \varepsilon_o E$ and $B = \mu_o H$, we have $p_A = p_M = S/c^2$, and no ambiguity surrounds the vacuum momentum of the fields. However, in material media, particularly those that possess large refractive indices, the two momentum densities are substantially different.

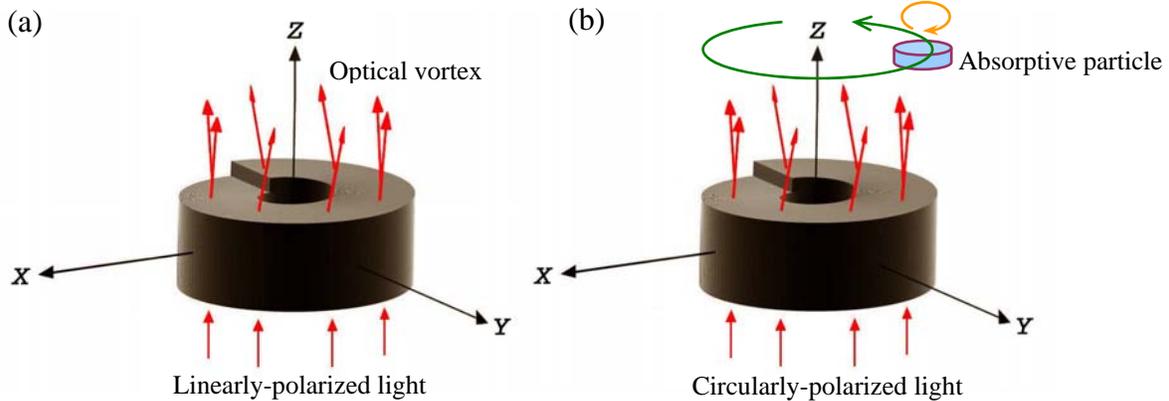

**Fig. 10**. (a) A transparent spiral ramp endows an incident beam with phase vorticity, which carries a certain amount of orbital angular momentum. (b) When the beam incident on the spiral ramp happens to be circularly polarized, the transmitted beam, a circularly-polarized optical vortex, carries both spin and orbital angular momenta. A small absorbing particle, placed in the path of such a beam, will spin on its own axis while, at the same time, travelling in a circle around the axis $z$ of the spiral ramp.

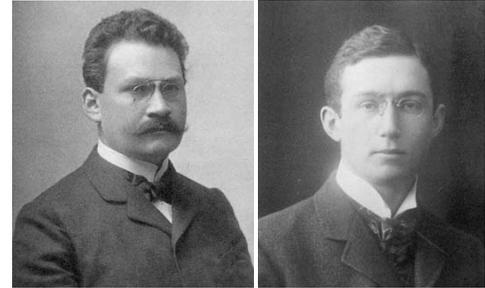

Hermann Minkowski (1864 – 1909)     Max Abraham (1875 – 1922)

As we have argued elsewhere [4,19], the controversy cannot be resolved within the confines of the classical theory, but requires arguments from without, which arguments must conform with logic, with physical law, and ultimately, of course, with experimental observations. One such argument which has been put forward by N. L. Balazs [33] will be described in Sec. 9. According to the Balazs thought experiment, the EM momentum density under the conditions of the thought experiment, is that of Abraham. We have generalized this result and postulated that, under any and all circumstances, the EM momentum density of the fields is $p_A(r,t) = S(r,t)/c^2$. A corollary of this postulate is that the field's angular momentum density is $L_A(r,t) = r \times S(r,t)/c^2$ [34].

In close association with the EM momentum, the atoms and molecules of material media experience forces and torques exerted upon them by the EM fields. It is through these forces and torques that linear and angular momenta are exchanged between the fields and the media that come into contact with the fields. The general expressions for EM force and torque, exerted by the $E$- and $H$-fields on the constituents of matter, namely, $\rho_{\text{free}}$, $J_{\text{free}}$, $P$ and $M$, are as follows [11,12,15-19,35]:

$$F(r,t) = \rho_{\text{free}} E + J_{\text{free}} \times \mu_o H + (P \cdot \nabla)E + (\partial P/\partial t) \times \mu_o H + (M \cdot \nabla)H - (\partial M/\partial t) \times \varepsilon_o E, \qquad (3)$$

$$T(r,t) = r \times F(r,t) + P(r,t) \times E(r,t) + M(r,t) \times H(r,t). \qquad (4)$$

The above expressions of force and torque, in conjunction with the earlier definitions of EM momentum densities (linear and angular) are fully consistent with the conservation laws. For instance, in a closed system such as that shown in Fig. 11, the total Abraham momentum at a given instant of time may be evaluated as $\boldsymbol{p}_{EM}(t) = \iiint (\boldsymbol{E} \times \boldsymbol{H}/c^2)\,\mathrm{d}\boldsymbol{r}$. Similarly, the total force exerted on the material media of the system may be obtained from Eq. (3) as $\boldsymbol{F}(t) = \iiint \boldsymbol{F}(\boldsymbol{r},t)\,\mathrm{d}\boldsymbol{r}$. It can then be shown under the most general conditions that $\boldsymbol{F}(t) = -\mathrm{d}\boldsymbol{p}_{EM}(t)/\mathrm{d}t$ [36]. A similar relation holds between the total torque exerted by the fields on the material media and the total EM angular momentum of the system. Not only are the force and torque expressions of Eqs. (3) and (4) consistent, in the context of momentum conservation, with the postulated linear and angular momentum densities, but also they are in complete accord with the laws of EM energy embodied in the Poynting vector and the associated formulas governing the exchange of energy between fields and matter [27].

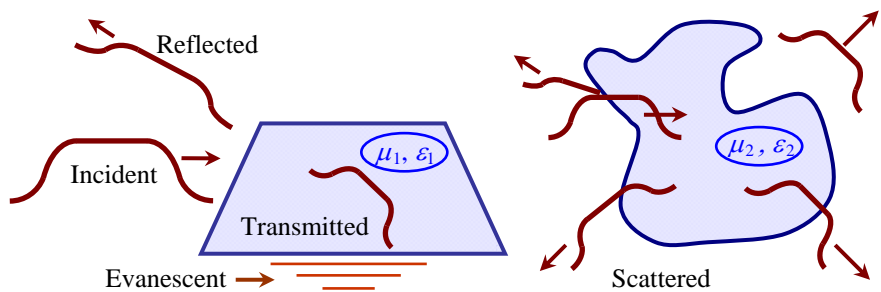

**Fig. 11**. In a closed system consisting of arbitrary material media and electromagnetic fields, the time rate of change of the total Abraham momentum is exactly equal in magnitude and opposite in direction to the total force exerted by the $E$ and $H$ fields on the media in accordance with Eq. (3).

It is often useful to examine the properties of a light pulse propagating in an isotropic and homogeneous dielectric of refractive index $n$. The pulse should be long enough that one could associate it with a single frequency $\omega_\mathrm{o}$, and broad and uniform enough over its cross-sectional area that one could readily associate it with a plane-wave. In other words, the pulse should have narrow spectra both in temporal and spatial frequency domains. The energy content of this pulse, $\mathcal{E}$, can be computed by integrating the corresponding Poynting vector $\boldsymbol{S}(\boldsymbol{r},t)$ over the pulse duration and over the cross-sectional area. Its electromagnetic (Abraham) momentum density can be similarly calculated by integrating $\boldsymbol{S}(\boldsymbol{r},t)/c^2$ over the pulse volume, and found to be $\mathcal{E}/(nc)$. The forces exerted on the host dielectric by the leading and trailing edges of the pulse are obtained from Eq. (3) as $F_z = \pm\frac{1}{4}\varepsilon_\mathrm{o}(n^2-1)E_\mathrm{o}^2$, where $F_z$ is the force per unit cross-sectional area in the direction of propagation, $z$, $\varepsilon_\mathrm{o}$ is the permittivity of free space, $n$ is the refractive index of the host medium, and $E_\mathrm{o}$ is the amplitude of the $E$-field at the center of the wavepacket [3]. The ± signs apply to the leading and trailing edges of the pulse, respectively, where the leading edge pushes the material forward while the trailing edge pulls it backward. These forces produce a mechanical momentum in the host medium that is essentially confined to the volume of the pulse and has a value of $\frac{1}{2}(n-n^{-1})\mathcal{E}/c$. The total momentum of the pulse (electromagnetic plus mechanical) is thus found to be $\frac{1}{2}(n+n^{-1})\mathcal{E}/c$, which is the arithmetic mean of the Abraham and Minkowski values.

Figure 12 shows one possible mechanism of entry into a dielectric medium for such a pulse. The reflectance and transmittance at the entrance facet being $R = (1-n)^2/(1+n)^2$ and $T = 4n/(1+n)^2$, it is clear that, if the incident pulse has energy $\mathcal{E}$ and momentum $\mathcal{E}/c$, then the reflected and transmitted pulses will have the momenta $-(1-n)^2 \mathcal{E}/[(1+n)^2 c]$ and $2n(n+n^{-1})\mathcal{E}/[(1+n)^2 c]$, respectively. The sum of these momenta is $\mathcal{E}/c$, indicating that the total momentum has the same value as it had before the pulse arrived at the entrance facet of the medium.

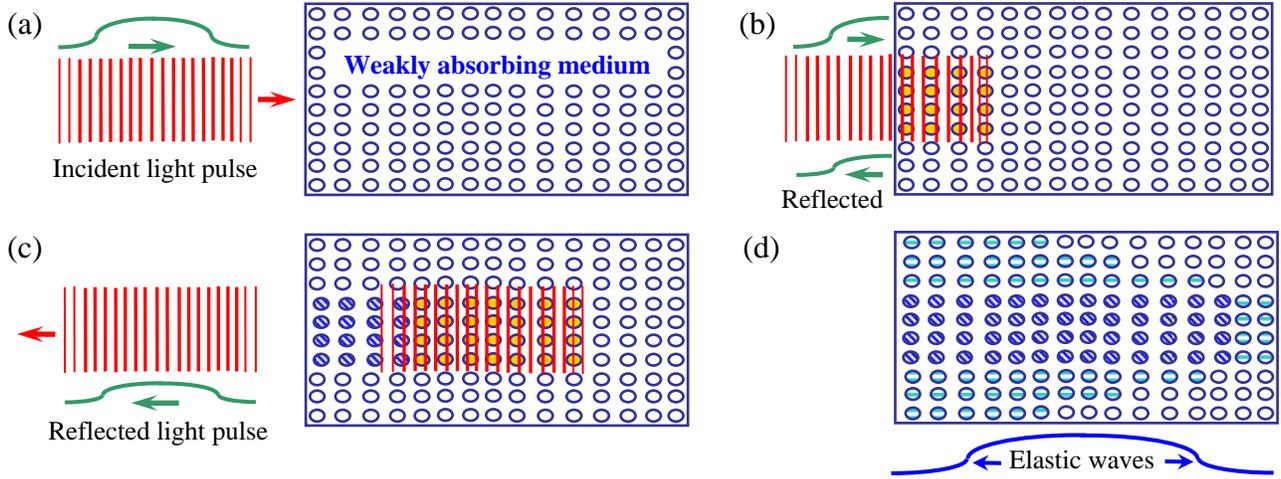

**Fig. 12**. Elastic waves launched within a weakly-absorbing medium upon entrance of a light pulse. The assumption of "weak absorption" is made here to avoid complications arising from the interaction between the light pulse and the rear facet of the slab upon exiting a transparent medium. The orange-colored atoms acquire mechanical momentum from the leading edge of the light pulse, relinquishing that momentum only after interacting with the trailing edge of the pulse. The forward-displaced atoms (dark blue) tend to drag their neighboring atoms (light blue) via elastic forces, and are restrained by the inertia of these neighbors.

**8. Photon momentum deduced from Fresnel coefficients at the vacuum-glass interface**. A simple argument advanced in the present section shows that the photon momentum inside a transparent dielectric must be equal to the arithmetic mean of the Abraham and Minkowski values. (The same conclusion was reached in the preceding section on the basis of the classical electrodynamics theory.)

Consider the glass slab of refractive index $n=1.5$ shown in Fig. 13. The Fresnel reflection coefficient at the entrance facet of the slab being $r = (1-n)/(1+n) = -0.2$, a total of 4% of all incident photons bounce back from the interface. Momentum conservation dictates that a reflected photon must transfer a momentum of $2\hbar\omega/c$ to the slab, while a transmitted photon must maintain its vacuum momentum of $\hbar\omega/c$. Assuming the incident pulse contains a total of 100 photons, the total momentum of the photons that enter the slab plus that of the slab itself should be $(96 + 4\times 2)\hbar\omega/c = 104\hbar\omega/c$. The momentum associated with individual photons that have entered the slab is then given by $(104/96)\hbar\omega/c = 1.0833\hbar\omega/c = 0.5(n + n^{-1})\hbar\omega/c$. This is precisely one-half the sum of Abraham and Minkowski values of the photon momentum inside the glass. The argument holds for any value of $n$ and any number of photons contained in the incident light pulse, provided, of course, that the number of incident photons is sufficiently large to justify statistical averaging. The same argument, when applied to the angular momentum of circularly-polarized photons, reveals the angular momentum of individual photons inside the dielectric to be the same as that in vacuum (i.e., $\hbar$), simply because reflected photons do not transfer any angular momentum to the glass slab upon reflection from its front facet.

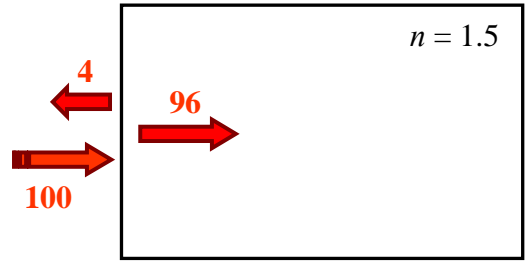

**Fig. 13**. Light pulse containing 100 photons arrives from free space at the entrance facet of a glass slab of refractive index $n=1.5$.

**9. The Balazs thought experiment**. In 1953, N. L. Balazs published a short paper [33] describing the following thought experiment. Consider a transparent dielectric (glass) rod of length $L$, refractive index $n$, and large mass $M$. Let a short light pulse enter the rod from the left and exit from the right, without losses due to absorption or scattering, or due to reflections at the entrance and exit facets. Balazs suggested three different schemes for avoiding reflection at the facet, but for our purposes it suffices to assume the existence of perfect antireflection coatings on these facets.

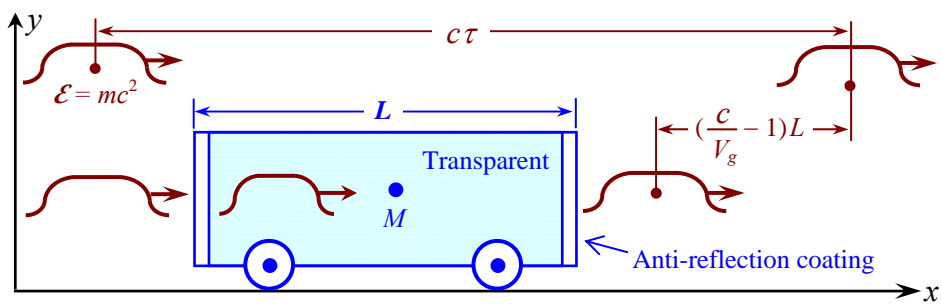

**Fig. 14**. The thought experiment of Balazs involves the propagation of a light pulse of energy $\mathcal{E}$ through a transparent rod of length $L$ and mass $M$. The rod can move on a frictionless rail along the $x$-axis. Since the group velocity $V_g$ of the pulse inside the rod is less than $c$, the emergent pulse is somewhat behind the location it would have reached had it travelled in vacuum all along.

When the pulse emerges from the rod it will be delayed by the reduced speed of light within the glass. In other words, had the pulse travelled parallel to its current path but outside the rod, it would have been ahead a distance of $(n-1)L$ compared to where it will be upon emerging from the rod. Since there are no external forces acting on the system of rod plus light pulse, the center-of-mass of the system should be in the same location irrespective of whether the pulse went through the rod or followed a parallel path outside the rod. Let the energy of the light pulse in vacuum be $\mathcal{E}$, which corresponds to a mass of $\mathcal{E}/c^2$. The delay has caused a leftward shift of the product of mass and displacement by $(n-1)L\mathcal{E}/c^2$. This must be compensated by a rightward shift of the rod itself. Let the light pulse have EM momentum $p$ while inside the rod. Considering that the momentum of the pulse before entering the rod is $\mathcal{E}/c$, the rod must have acquired a net momentum of $(\mathcal{E}/c) - p$ while the pulse travelled inside. Its net mass times forward displacement, therefore, must be $[(\mathcal{E}/c) - p]nL/c$. Equating the rightward and leftward mass × displacements yields $p = \mathcal{E}/(nc)$ for the EM momentum of the pulse inside the rod. This argument not only assigns the Abraham value to the EM momentum of the light pulse, but also indicates that the refractive index appearing in the Abraham expression is the group index (as opposed to the phase index) of the transparent medium.

In the remainder of this section we present a quantum interpretation of the Balazs thought experiment, which, although leading to the same conclusions, offers a fresh perspective on the problem. Instead of a light pulse of energy $\mathcal{E}$, suppose a single photon of energy $\hbar\omega$ arrives at the left facet of the rod. The rod will be assumed to have the same (group) refractive index $n$ as before, but its entrance and exit facets are no longer pacified with antireflection coatings. Upon each encounter with the facets, the photon will do one of two things: it either gets reflected at the interface or goes through. Therefore, there will be an infinite number of possible outcomes for the experiment: The photon might bounce back from the first facet, it might go through both facets successively (without any reflections at all), it might enter through the first facet, bounce off the second facet, and emerge from the rod at the left facet, etc. Each of these infinite possibilities has a finite probability of occurrence. An important observation is that, by placing photodetectors on both sides of the rod and by monitoring the delay between the time that the photon leaves its source and the time that it gets detected (either on the right- or on the left-side of the rod) it is possible to know whether the photon was transmitted or reflected, and also how many bounces it suffered within the rod before re-emerging into free space. This post-selection procedure allows one to exploit momentum conservation in order to determine the photon momentum inside its dielectric host.

The simplest possibility is for the photon to bounce back upon its first encounter with the glass rod. Its free-space momentum of $\hbar\omega/c$ will then reverse direction, and the rod will acquire a forward momentum of $2\hbar\omega/c$. Clearly the center-of-mass of the system would continue to move forward at the same rate as it would have, had the light pulse been travelling outside the rod. We do not learn anything about the EM momentum inside the rod, because the photon never enters the glass medium. The next possibility is for the photon to enter the rod, propagate the length $L$ of the rod, then exit from the second facet. Denoting the photon's EM momentum inside the glass by $p$, Conservation of momentum throughout the process shows that the rod will acquire a momentum of $[(\hbar\omega/c) - p]$ while the photon is inside. Equating the rightward mass × displacement of the rod, $[(\hbar\omega/c) - p]nL/c$, with the leftward mass × displacement for the photon, $(n-1)L\hbar\omega/c^2$, yields $p = \hbar\omega/nc$, which is the same as in the original thought experiment.

Other possibilities can be treated in a similar way, and the final result will end up being the same as before. For instance, if the photon enters the rod from the left, turns around twice inside the rod, then emerges on the right-hand-side, it will have spent a total of $3nL/c$ seconds inside the rod. Upon emerging on the right-hand-side, the photon will have fallen behind by $(3n-1)L$ relative to its position had it been traveling in free space all along. The leftward mass × displacement of the photon will thus be $(3n-1)L\hbar\omega/c^2$. The rightward displacement of the rod now consists of three segments. In the first and third segments, the photon has been traveling to the right, and the mechanical momentum of the rod must have been $[(\hbar\omega/c) - p]$. The mass × displacement for each of these segments is thus going to be $[(\hbar\omega/c) - p]nL/c$. In the second segment, however, the photon travels to the left, and the corresponding mass × displacement of the rod is given by $[(\hbar\omega/c) + p]nL/c$. The total rightward mass × displacement of the rod is then $[3(\hbar\omega/c) - p]nL/c$, which, when equated with the corresponding entity for the photon would yield $p = \hbar\omega/nc$. The EM momentum of the photon inside glass is thus going to have the Abraham value under all circumstances.

**10. Photon absorbed by a submerged atom**. In a recent article [37], S. Barnett has cited the example of absorption of a photon by an atom submerged in a dielectric liquid [38] in support of his contention that the Minkowski momentum is the photon's canonical momentum. The argument, in a slightly simplified form, may be summarized as follows. Let the atom have a rest mass $m$ and a transition energy tuned to absorb a photon of frequency $\omega$ traveling in its homogeneous liquid host of refractive index $n$. Before absorption the atom is stationary, while afterward it acquires a velocity $V$ in the direction of propagation of the photon. Thus, in the rest-frame of the atom, the photon appears to have been Doppler-shifted to a slightly lower frequency, $\omega' = \omega(1 - \tfrac{1}{2}nV/c)$, with $\tfrac{1}{2}V$ being the average velocity of the atom during the absorption process. Before absorption, the energy of the system consisting of the photon, the atom, and the immersion liquid is $\hbar\omega$, whereas afterward the energy is $\tfrac{1}{2}mV^2+\hbar\omega'$. (In this analysis, it is permissible to ignore the kinetic energy of the liquid molecules that are dragged along by the motion of the atom, as the corresponding mass of these molecules may be assumed to have been included in the effective mass $m$ of the atom.) Conservation of energy before and after absorption now yields $\hbar\omega\approx\tfrac{1}{2}mV^2+\hbar\omega(1-\tfrac{1}{2}nV/c)$, which leads to $mV\approx n\hbar\omega/c$. According to this argument, the momentum $mV$ picked up by the atom, which must be the same as the photon momentum inside the liquid, turns out to be the Minkowski momentum.

The problem with the above argument is the neglect of the mechanical energy needed to drag the liquid molecules behind the atom, which molecules are subject to a Lorentz force exerted by the trailing edge of the photon wavepacket. The force per unit cross-sectional area at the trailing edge of the packet is $F_z=-\tfrac{1}{4}\varepsilon_o(n^2-1)E_o^2$, where $\varepsilon_o$ is the permittivity of free space, $n$ is the refractive index of the liquid, and $E_o$ is the $E$-field amplitude at the center of the packet [3]. Assuming a cross-sectional area $A$ and a duration $T$ for the photon, the packet's energy content is $\tfrac{1}{2}nE_o^2AT/Z_o$, where $Z_o=\sqrt{\mu_o/\varepsilon_o}$ is the impedance of free space. Equating this photon energy with $\hbar\omega$ yields $E_o^2AT=2Z_o\hbar\omega/n$. Now, the liquid molecules under the influence of the Lorentz force of the trailing edge travel an average distance of $\tfrac{1}{2}VT$ while the photon is being absorbed. The energy needed to drag these molecules is $\tfrac{1}{2}VTA|F_z|=\tfrac{1}{4}(n-n^{-1})V\hbar\omega/c$, which, when included in the preceding energy balance equation, yields $mV\approx\tfrac{1}{2}(n+n^{-1})\hbar\omega/c$. The photon momentum in the liquid is thus seen to be the arithmetic mean of the Abraham and Minkowski momenta, in agreement with numerous other derivations [3,4].

**11. Photon momentum deduced from the uncertainty principle**. An argument has been advanced by M. Padgett with regard to the momentum of photons inside dielectric media [39]. The original argument invokes the diffraction of monochromatic light upon passage through a narrow slit in the wall of a chamber filled with a transparent liquid, and the subsequent diffraction broadening of the transmitted light within the liquid. The argument can be made equivalently with reference to the system depicted in Fig. 15, where a monochromatic plane-wave is focused via a diffraction-limited lens into a hemispherical dielectric of refractive index $n$. The focused spot diameter at the focal plane of the lens, which coincides with the base of the hemisphere, is $\sim\lambda_o/(n\sin\theta)$, where $\theta$ is the half-angle of the focused cone. In the absence of the hemisphere, of course, the spot size will be $\lambda_o/\sin\theta$.

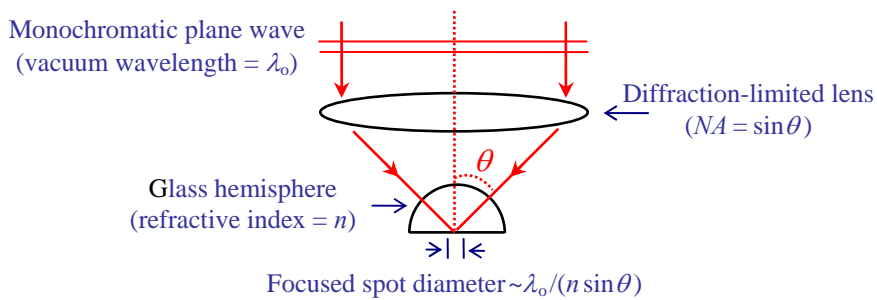

**Fig. 15**. The uncertainty in the photon's lateral position at the base of the glass hemisphere of refractive index $n$ is inversely proportional to $n$. Heisenberg's uncertainty principle thus dictates that the Minkowski momentum be associated with the photon inside the hemisphere.

If the incident beam is assumed to contain a single photon, its propagation through the medium of refractive index $n$ has reduced the uncertainty of the photon's lateral position at the base of the hemisphere by a factor of *n*. Heisenberg's uncertainty principle then implies that the momentum uncertainty within the hemisphere must be *n* times greater than that in vacuum and that, therefore, the photon momentum inside the dielectric must coincide with the Minkowski momentum. This is obviously a purely quantum-mechanical argument, as it involves the uncertainty principle; no classical reasoning can be advanced to confirm or to refute its conclusion. However, since the uncertainty principle is a relation between conjugate variables, the canonical commutation relation $[x, p_x] = i\hbar$ between the coordinate *x* and its conjugate momentum $p_x$ implies that the photon momentum in Padgett's argument is the canonical momentum. Barnett's contention that the Minkowski momentum of a photon is its canonical momentum [37] is, therefore, in complete accord with Padgett's reasoning. None of this, however, sheds any light on the kinetic momentum of the photon, which, from every other consideration, appears to have the Abraham value.

**12. Concluding Remarks**. We have shown that the exchange of energy, linear momentum, and angular momentum between electromagnetic waves and material media can have significant scientific as well as technological applications, particularly in biological research and in the areas of micro/nano-scale motors, actuators, and manipulators. The generalized Lorentz laws of force and torque given by Eqs. (3) and (4), in conjunction with Maxwell's macroscopic equations and the constitutive relations, give a consistent explanation of the momentum of light and radiation pressure on material media. The formulation is quite general and has been incorporated into a Finite Difference Time Domain (FDTD) Maxwell-solver to yield the distributions of force and torque in diverse situations of practical interest [40,41]. It is the author's belief that the long-standing Abraham-Minkowski controversy surrounding the momentum of light in material media has finally been resolved in favor of the Abraham momentum.

**Dedication**. This paper is dedicated to my PhD dissertation advisor, Professor Joseph W. Goodman, who has been an inspiration and a role model throughout my career.

**References**

1. R. P. Feynman, R. B. Leighton, and M. Sands, *The Feynman Lectures on Physics*, Vol. II, Chap. 27, Addison-Wesley, Reading, Massachusetts (1964).
2. J. D. Jackson, *Classical Electrodynamics*, 2nd edition, Wiley, New York, 1975.
3. M. Mansuripur, "Radiation pressure and the linear momentum of the electromagnetic field," *Optics Express* **12**, 5375-5401 (2004).
4. M. Mansuripur, "Resolution of the Abraham-Minkowski controversy," Optics Communication **283**, 1997-2005 (2010).
5. M. Mansuripur, A. R. Zakharian, and J. V. Moloney, "Radiation pressure on a dielectric wedge," Optics Express **13**, 2064-2074 (2005).
6. A. Ashkin, *Phys. Rev. Lett.* **24,** 156 (1970).
7. A. Ashkin and J. M. Dziedzic, *Phys. Rev. Lett.* **30**, 139 (1973).
8. A. Ashkin, J. M. Dziedzic, J. E. Bjorkholm and S. Chu, "Observation of a single-beam gradient force optical trap for dielectric particles," *Opt. Lett.* **11**, 288-290 (1986).
9. A. Ashkin and J. M. Dziedzic, "Optical trapping and manipulation of viruses and bacteria," *Science* **235**, 1517-1520 (1987).
10. M. Mansuripur, "Spin and orbital angular momenta of electromagnetic waves in free space," to appear in Phys. Rev. A, 2011.
11. A. Einstein and J. Laub, "Über die elektromagnetischen Grundgleichungen für bewegte Körper," *Annalen der Physik* **331**, 532–540 (1908).

12. A. Einstein and J. Laub, "Über die im elektromagnetischen Felde auf ruhende Körper ausgeübten ponderomotorischen Kräfte," *Annalen der Physik* **331**, 541–550 (1908).
13. R. Peierls, "The momentum of light in a refracting medium," Proc. Roy. Soc. Lond. A. **347**, 475-491 (1976).
14. R. Loudon, L. Allen, and D. F. Nelson, "Propagation of electromagnetic energy and momentum through an absorbing dielectric," Phys. Rev. E **55**, 1071-1085 (1997).
15. M. Mansuripur, "Radiation pressure and the linear momentum of the electromagnetic field in magnetic media," *Optics Express* **15**, 13502-18 (2007).
16. M. Mansuripur, "Electromagnetic Stress Tensor in Ponderable Media," *Optics Express* **16**, 5193-98 (2008).
17. M. Mansuripur, "Electromagnetic force and torque in ponderable media," *Optics Express* **16**, 14821-35 (2008).
18. M. Mansuripur, "Generalized Lorentz law and the force of radiation on magnetic dielectrics," *SPIE Proc.* **7038,** 70381T (2008).
19. M. Mansuripur and A. R. Zakharian, "Maxwell's macroscopic equations, the energy-momentum postulates, and the Lorentz law of force," *Phys. Rev. E* **79**, 026608 (2009).
20. L. Allen, M. W. Beijersbergen, R. J. C. Spreeuw, and J. P. Woerdman, "Orbital angular-momentum of light and the transformation of Laguerre-Gaussian laser modes," Phys. Rev. A **45**, 8185-8189 (1992).
21. N. B. Simpson, K. Dholakia, L. Allen, and M. J. Padgett, "Mechanical equivalence of spin and orbital angular momentum of light: An optical spanner," Opt. Lett. **22**, 52-54 (1997).
22. A. T. O'Neil, I. MacVicar, L. Allen, and M. J. Padgett, "Intrinsic and Extrinsic Nature of the Orbital Angular Momentum of a Light Beam," Phys. Rev. Lett. **88**, 053601 (2002).
23. S. M. Barnett, "Rotation of electromagnetic fields and the nature of optical angular momentum," J. Mod. Opt. **57**, 1339-1343 (2010).
24. L. Marrucci, E. Karimi, S. Slussarenko, B. Piccirillo, E. Santamato, E. Nagali, and F. Sciarrino, "Spin-to-orbital conversion of the angular momentum of light and its classical and quantum applications," J. Opt. **13**, 064001 (2011).
25. T. A. Nieminen, T. Asavei, V. L. Y. Loke, N. R. Heckenberg, and H. Rubinsztein-Dunlop, "Symmetry and the generation and measurement of optical torque," J. Quant. Spect. Rad. Trans. **110**, 1472-1482 (2009).
26. L. Allen, M. J. Padgett, and M. Babiker, "The orbital angular momentum of light," Prog. Opt. **39**, 291-372 (1999).
27. M. Mansuripur, *Field, Force, Energy and Momentum in Classical Electrodynamics*, Bentham e-books, Chap.10 (2011).
28. M. E. J. Friese, H. Rubinsztein-Dunlop, J. Gold, P. Hagberg, and D. Hanstorp, "Optically driven micromachine elements," Appl. Phys. Lett. **78**, 547-549 (2001).
29. D. G. Grier, "A revolution in optical manipulation," Nature **24**, 810-816 (2003).
30. R. Loudon, "Radiation Pressure and Momentum in Dielectrics," De Martini lecture, Fortschr. Phys. **52**, 1134-40 (2004).
31. S. M. Barnett and R. Loudon, "On the electromagnetic force on a dielectric medium," *J. Phys. B: At. Mol. Opt. Phys.* **39**, S671-S684 (2006).
32. S. M. Barnett, R. Loudon, "The enigma of optical momentum in a medium," Phil. Trans. Roy. Soc. A **368**, 927-39 (2010).
33. N. L. Balazs, "The energy-momentum tensor of the electromagnetic field inside matter," Phys. Rev. **91**, 408-411 (1953).
34. M. Mansuripur, "Nature of electric and magnetic dipoles gleaned from the Poynting theorem and the Lorentz force law of classical electrodynamics," Optics Communications **284**, 594-602 (2011).
35. R. M. Fano, L. J. Chu, and R. B. Adler, *Electromagnetic Fields, Energy and Forces*, Wiley, New York (1960).
36. M. Mansuripur and A. R. Zakharian, "Energy, momentum, and force in classical electrodynamics: application to negative-index media," Optics Communications **283**, 4594-4600 (2010).
37. S. M. Barnett, "Resolution of the Abraham-Minkowski Dilemma," Phys. Rev. Lett. **104**, 070401 (2010).
38. D. H. Bradshaw, Z. Shi, R. W. Boyd, and P. W. Milonni, "Electromagnetic momenta and forces in dispersive dielectric media," Optics Communications **283**, 650 (2010).
39. M. J. Padgett, "On diffraction within a dielectric medium as an example of the Minkowski formulation of optical momentum", Optics Express **16**, 20864-68 (2008).
40. A. R. Zakharian, M. Mansuripur, and J. V. Moloney, "Radiation pressure and the distribution of electro-magnetic force in dielectric media," *Optics Express* **13**, 2321-36 (2005).
41. M. Mansuripur, A. R. Zakharian, "Whence the Minkowski momentum?" Optics Communications **283**, 3557-63 (2010).